\documentclass[aps,pre,reprint,longbibliography,nobalancelastpage,superscriptaddress,floatfix]{revtex4-2}

\usepackage{amsmath,amssymb,amsthm,amsbsy}
\usepackage[utf8]{inputenc}
\usepackage[T1]{fontenc}
\usepackage{graphicx}
\usepackage{xcolor}
\usepackage{mathrsfs}
\usepackage{stmaryrd}
\usepackage{hyperref}
\usepackage{enumitem}
\usepackage{accents}
\usepackage{newtxtext}
\usepackage[slantedGreek]{newtxmath}
\usepackage{tikz}
\usepackage[compact]{titlesec}

\usepackage{multirow}

\usetikzlibrary{tikzmark}

\definecolor{linkcolor}{HTML}{223096}
\hypersetup{colorlinks,allcolors=linkcolor}
\bibpunct{\textcolor{linkcolor}{[}}{\textcolor{linkcolor}{]}}{\textcolor{linkcolor}{,}}{n}{}{;}
\renewcommand{\eqref}[1]{\hyperref[#1]{(\ref*{#1})}}

\usetikzlibrary{decorations.pathreplacing,arrows}

\renewcommand{\vec}[1]{\boldsymbol{#1}}

\newcommand{\btilde}[1]{\boldsymbol{\tilde{#1}}}

\DeclareMathAlphabet{\mathcal}{OMS}{cmsy}{m}{n}
\DeclareMathAlphabet{\mathbfsf}{\encodingdefault}{\sfdefault}{b}{n}

\newcommand{\tens}[1]{\mathbfsf{#1}}

\newcommand{\figref}[2]{[Fig.~\hyperref[#1]{\ref*{#1}(#2)}]}
\newcommand{\bfigref}[3]{[Fig.~\hyperref[#1]{\ref*{#1}(#2\textsubscript{#3})}]}
\newcommand{\figrefi}[2]{[Fig.~\hyperref[#1]{\ref*{#1}(#2)}, inset]}
\newcommand{\textfigref}[2]{Fig.~\hyperref[#1]{\ref*{#1}(#2)}}
\newcommand{\textfigureref}[2]{Figure~\hyperref[#1]{\ref*{#1}(#2)}}
\newcommand{\wholefigref}[1]{(Fig.~\ref{#1})}
\newcommand{\wholefigrefi}[1]{(Fig.~\ref{#1}, inset)}
\newcommand{\textwholefigref}[1]{Fig.~\ref{#1}}

\newcommand{\figrefp}[2]{\hyperref[#1]{\ref*{#1}(#2)}}

\newcommand{\citeappendix}[1]{Appendix~\ref{#1}}
\renewcommand{\leq}{\leqslant}

\DeclareMathAlphabet{\mathcal}{OMS}{cmsy}{m}{n}

\renewcommand{\citet}[1]{{\protect\NoHyper\citeauthor{#1}~(\citeyear{#1})} \cite{#1}}

\renewcommand{\pi}{\uppi}
\renewcommand{\Delta}{\upDelta}

\begin{document}

\title{A mechanical bifurcation constrains the evolution of cell sheet folding in the family Volvocaceae}
\author{Valens Tribet}
\affiliation{D\'epartement de Physique, \'Ecole Normale Sup\'erieure, 24 rue Lhomond, 75005 Paris, France}
\affiliation{Max Planck Institute for the Physics of Complex Systems, N\"othnitzer Stra\ss e 38, 01187 Dresden, Germany}
\affiliation{Center for Systems Biology Dresden, Pfotenhauerstra\ss e 108, 01307 Dresden, Germany}
\author{Pierre A. Haas}
\email[Contact author: ]{haas@pks.mpg.de}
\affiliation{Max Planck Institute for the Physics of Complex Systems, N\"othnitzer Stra\ss e 38, 01187 Dresden, Germany}
\affiliation{Center for Systems Biology Dresden, Pfotenhauerstra\ss e 108, 01307 Dresden, Germany}
\affiliation{\smash{Max Planck Institute of Molecular Cell Biology and Genetics, Pfotenhauerstra\ss e 108, 01307 Dresden, Germany}}
\date{\today}%
\begin{abstract}
The processes of morphogenesis that give rise to the shapes of organs and organisms during development are often driven by mechanical instabilities. Can such mechanical bifurcations also drive or constrain the evolution of these processes in the first place? We discover an instance of these constraints in the green algae of the family Volvocaceae. During their development, their bowl-shaped embryonic cell sheet turns itself inside out. This inversion is driven by a simple wave of cell wedging in the genus \emph{Pleodorina} (16--128 cells) and more complex programmes of cell shape changes in \emph{Volvox} ($\sim$400--50\,000 cells). However, no species with intermediate cell numbers (256 cells) have been described. Here, we relate this gap to a mechanical bifurcation: Focusing on the inversion of \emph{Pleodorina californica} (64 cells), we develop a continuum model, in which the cell shape changes driving inversion appear as changes of the intrinsic curvature of an elastic surface. A mechanical bifurcation in this model predicts that inversion is only possible in a subset of its parameter space. Strikingly, parameters estimated for \emph{P.~californica} fall into this possible subset, but those that we extrapolate to 256 or more cells using allometric observations and a model of cell cleavage in Volvocaceae do not. Our work thus suggests that the more complex inversion strategies of \emph{Volvox} are an evolutionary necessity to obviate this bifurcation and indicates more broadly how mechanical bifurcations can drive the evolution of morphogenesis.
\end{abstract}

\maketitle
\section{\uppercase{Introduction}}
Evolution has not only led to the shapes of tissues, organs, and organisms, but also to the morphogenetic processes through which these shapes arise during development. The emergence of these processes need not however be driven only by evolutionary pressures, but can also be driven by the physics and mechanics of morphogenesis. The cephalic furrow~\cite{vellutini23,dey23}, a tissue fold that appears transiently during \emph{Drosophila} gastrulation, provides a recent example of this: in mutants in which the cephalic furrow does not form, ectopic folds appear, suggesting that the cephalic furrow evolved as a mechanical means to absorb compressive stresses associated with germband extension~\cite{vellutini23}. Interestingly, a different strategy to reduce compressive stresses (viz., out-of-plane cell divisions) has evolved in other fly species that lack a cephalic furrow~\cite{dey23}.

Over the last decade or so, mechanical instabilities~\cite{nelson16} have been recognised as drivers of the emergence of shapes in diverse developmental processes~\cite{savin11,tallinen14,varner15,engstrom18,montandon19,gill24,alber25}. This prompts the question: Can such mechanical bifurcations also constrain or limit the evolution of the mechanical processes of morphogenesis? 

Here, we discover an instance of these mechanical limits, by showing how a mechanical bifurcation constrains the evolution of cell sheet folding in the multicellular colonial algae of the family Volvocaceae that we introduce in detail below. The embryonic cell sheet of the Volvocaceae is bowl-shaped or spherical, and turns itself inside out in a process called inversion \mbox{\cite{kirkreview,hallmann2006morphogenesis,matt16}}. In the genus \emph{Pleodorina} [16--128 cells, \textfigref{fig1}{a}], this inversion is driven by a wave of cell wedging~\cite{hohn2016distinct}, but more complex inversion processes~\cite{hallmann2006morphogenesis} have evolved in the larger species of \emph{Volvox} [$\sim$400--50\,000 cells, \textfigref{fig1}{a}]. We develop a continuum model of the inversion of \emph{Pleodorina californica} (64 cells), in which the cell sheet is an elastic surface to which the cell shape changes driving inversion impart changes of intrinsic curvature. In this model, we find a mechanical bifurcation that implies that inversion is only possible in a subset of its parameter space. Our parameter estimates for \emph{P.~californica} fall into this range. Moreover, we discover allometric scalings of the model parameters across Volvocaceae. Together with a model of cell division, these allow us to show by extrapolation that hypothetical species with 256 cells do not fall into this parameter range. This indicates that the fact that no such species have been described is a consequence of this bifurcation and, in turn, that the morphogenetic innovations of \emph{Volvox} were an evolutionary requirement arising from the bifurcation. Our work thus showcases how the evolution of morphogenesis can be constrained by mechanical bifurcations.

\begin{figure}
\centering\includegraphics{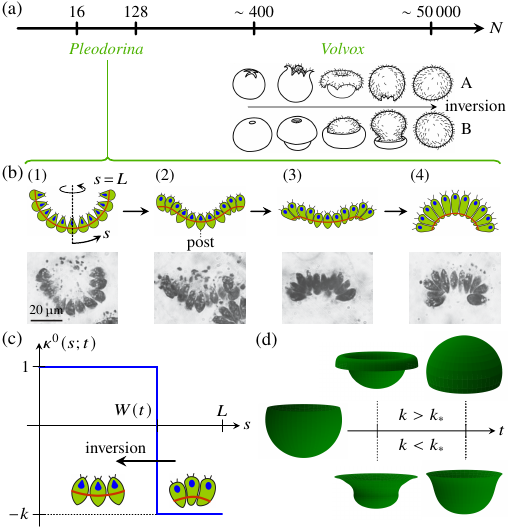}
\caption{Inversion in \emph{Pleodorina californica}: mechanical model and bifurcation. (a) Cell numbers $N$ of the genera \emph{Pleodorina} and \emph{Volvox} in the family Volvocaceae. Inset: Schematics of type-A and type-B inversion in different species of \emph{Volvox}, reproduced from Ref.~\cite{hohn2015inversion}. (b)~Inversion of \emph{P.~californica}, reproduced from Ref.~\cite{hohn2016distinct}. Top row: schematics of a section of the axisymmetric bowl-shaped cell sheet at different stages of inversion; the length of the cross-section before inversion is $L$, and arclength along the cross-section is $s$. Red line: position of cytoplasmic bridges; post: posterior pole. Bottom row: corresponding light micrographs of stained semi-thin sections. Scale bar: $20\,\text{\textmu m}$. (c)~Mechanical model of inversion in \emph{P.~californica}. The cell sheet is represented as an elastic spherical cap of radius $R=1$. The slow cell shape changes of inversion are encoded in quasistatic variations of its intrinsic curvature $\kappa^0(s;t)$, where $s\in [0,L]$ is undeformed arclength, and $t$ is the time of inversion. The front of the wave of cell shape changes at $s=W(t)$ separates the uninverted part of the cell sheet (where cells are spindle-shaped, $\kappa^0=1$) from the inverted part (where cells are more wedge-shaped, $\kappa^0=-k<0$). As inversion progresses, $W(t)$ decreases from $s=L$ towards $s=0$ (arrow). Insets: cell shapes on either side of the front, from Ref.~\cite{hohn2016distinct}. (d) Mechanical bifurcation: If $k>k_\ast$ is large enough (top), the edges of the cap flip over as the bend region progresses and the cell sheet inverts (shape insets at different times~$t$); if $k<k_\ast$ is too small, the edges do not flip over and inversion fails.}\label{fig1} 
\end{figure}

\subsection*{Inversion in the family Volvocaceae}\label{ssec:inversion bio}
The family Volvocaceae is a model for the evolution of multicellularity~\cite{kirkessay,herron16} and, together with the closely related single-celled alga \emph{Chlamydomonas}, has also become a model for biological fluid mechanics~\cite{goldstein15}. More recently, the Volvocaceae have begun to emerge as models for cell sheet folding in development because of the fascinating process of inversion~\citep{kirkreview,hallmann2006morphogenesis,matt16} by which the bowl-shaped or spherical embryonic cell sheet turns itself inside out at the end of cell division. Through inversion, those cells poles whence will emanate cilia are exposed to the outside to enable motility.

In the largest species, which are found in the genus \emph{Volvox}, inversion proceeds in at least two different, species-specific ways~\cite{hallmann2006morphogenesis}: In type-A inversion \cite{viamontes1977cell,viamontes79,green81,haas2018embryonic}, four lips flip over at the anterior pole of the cell sheet to pull it back and hence invert it \figrefi{fig1}{a}. In type-B inversion \citep{hohn11,hohn2015inversion,haas2018noisy,haas25}, a circular invagination at the equator enables inversion of the posterior hemisphere; slightly later, an opening at the top of the cell sheet widens and the anterior hemisphere inverts by peeling over the inverted posterior \figrefi{fig1}{a}. Other genera within Volvocaceae also undergo inversion \cite{hallmann2006morphogenesis}, but have been studied, in general, less thoroughly than the genus \emph{Volvox}. One notable exception is \emph{Pleodorina californica}~\figref{fig1}{b}, which inverts by a wave of cell shape changes travelling from the edge of the hemispherical cell sheet to its posterior pole \cite{hohn2016distinct}, in a manner reminiscent of type-A inversion in \emph{Volvox} but lacking the lips. Before inversion, the cells are teardrop-shaped~\bfigref{fig1}{b}{1}. At the start of inversion~\bfigref{fig1}{b}{2}, cells become spindle-shaped. Wedge-shaped cells appear at the edge of the cell sheet: the apical poles of these (i.e., the ends from which cilia will grow) become rounder, while their basal poles thin. This inverts the ``polarity'' of the initial teardrop shapes. As inversion proceeds~\bfigref{fig1}{b}{3}, this cell shape change propagates in a wave to the posterior pole of the cell sheet to drive complete inversion of its curvature~\bfigref{fig1}{b}{4}. The cells are connected to their neighbours by cytoplasmic bridges~\cite{hohn2016distinct}, and these cell shape changes are accompanied by migration of cytoplasmic bridges from their midplanes to their basal poles. This migration splays the cells and hence bends the cell sheet. Analogous motion of cytoplasmic bridges is known to be required for type-A inversion in \emph{V.~carteri}~\cite{nishii03}. 

\section{\uppercase{Results}}
\subsection{Mechanical model of inversion in \emph{P.~californica}}
We begin by setting up a mechanical model of inversion in \emph{P.~californica}, in which the cell sheet is an elastic sheet and the cell shape changes driving inversion appear as changes of the intrinsic, preferred geometry of this sheet. This model is in the spirit of previous mechanical models of inversion in \emph{Volvox} \cite{hohn2015inversion,haas15,haas2018noisy,haas2018embryonic,haas25}. The deformations of the cell sheet during inversion are approximately axisymmetric. We therefore describe the cell sheet by its undeformed arclength $s$ from the posterior pole at $s=0$~\bfigref{fig1}{b}{1}. The undeformed configuration of the cell sheet is a spherical cap of radius $R=1$ and thickness $h$, with $s=L$ corresponding to the rim of the cell sheet~\bfigref{fig1}{b}{1}. We denote by $\kappa^0$ the intrinsic curvature of the cell sheet.

Close to the posterior pole, the cells are spindle-shaped during inversion~\bfigref{fig1}{b}{2}. These spindle shapes, illustrated in Fig.~10(b) of Ref.~\cite{hohn2016distinct}, favour the uninverted curvature of the cell sheet, so that $\kappa^0=1$ there. We emphasise that the identically named spindle-shaped cells that appear during \mbox{type-B} inversion in \emph{V.~globator}, shown in Fig.~12(b) of Ref.~\cite{hohn11}, have different shapes: They are symmetric across the plane of cytoplasmic bridges, which suggests that $\kappa^0=0$ for those cells. This mechanical hypothesis has been verified in Ref.~\cite{haas25}. By contrast, $\kappa^0=-k<0$ in the region of elongated cells behind the front of cell shape changes~\bfigref{fig1}{b}{3}. The fact that the radius of the cell sheet is smaller post-inversion than its pre-inversion radius~\figref{fig1}{b} suggests that $k>1$. Indeed, the cell size measurements of Ref.~\cite{hohn2016distinct} indicate that the cell shape changes of \emph{P.~californica}, unlike the cell shape changes of \emph{Volvox}~\cite{viamontes79,hohn11,hohn2015inversion,haas2018embryonic}, do not involve contraction or expansion within the plane of the cell sheet. The intrinsic curvature of the cell sheet is therefore~\figref{fig1}{c}
\begin{align}
\kappa^0(s;t) = \left\{\begin{array}{cl}
1&\text{if }0<s<W(t),\\
-k&\text{if }W(t)<s<L,\\
\end{array}\right.\label{eq:k0}
\end{align}
where $t$ is the time of inversion, and $W(t)$ is the position of the front of the wave of cell shape changes. During inversion, $W(t)$ decreases quasi-statically from $s=L$ towards $s=0$, and the deformed shape of the cell sheet minimises its elastic energy $\mathcal{E}$ instantaneously. The form of the elastic energy results from a morphoelastic shell theory for large bending deformations~\citep{haas2021morphoelasticity}, which we describe in detail in \citeappendix{appA}. 

\begin{figure}[t!]
\includegraphics{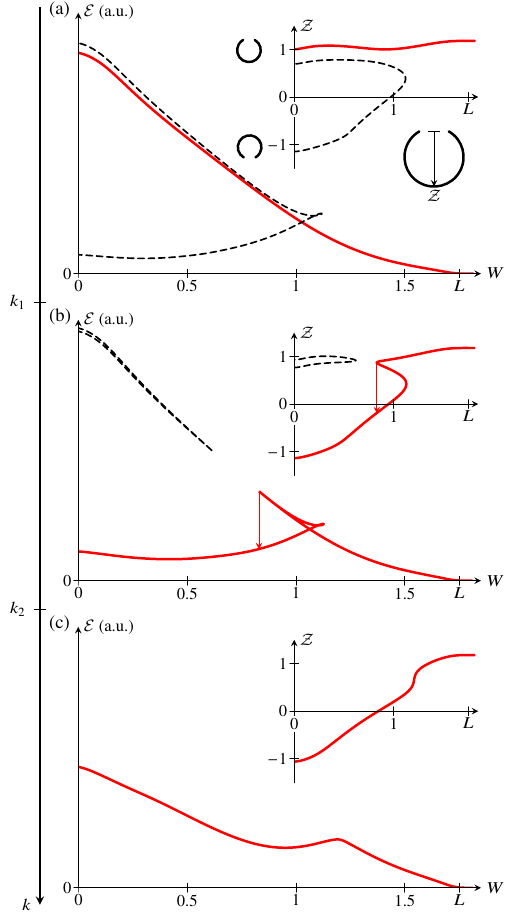}
\centering
\caption{Bifurcation diagrams for inversion in \emph{P.~californica}. Plots of the elastic energy $\mathcal{E}$ against the position $W$ of the front of the wave of cell shape changes, for different values of the absolute value $k$ of the inverted intrinsic curvature. Insets: corresponding plots of the signed distance $\mathcal{Z}$ [defined in panel (a)]; $\mathcal{Z}>0$ ($\mathcal{Z}<0$) indicates inverted (uninverted) configurations of the cell sheet. Bifurcations at $k=k_1$ and $k=k_2$ separate three qualitatively different bifurcation diagrams. (a)~For $k<k_1$, the cell sheet does not invert on the branch of lowest energy (red line) connected to the undeformed configuration. The cell sheet does invert on another branch (dashed line) of lower energy, but this branch is not connected to the undeformed configuration. (b)~For $k_1<k<k_2$, the cell sheet inverts on the branch of lowest energy (red line). Fold points on this branch imply a discontinuous snapthrough as $W$ decreases (arrows). The branch on which the lips do not flip over (dashed line) is no longer connected to the undeformed configuration. (c)~For $k>k_2$, the cell sheet inverts without snapthrough on the branch of lowest energy. Parameter values: $L=1.75$, $h=0.15$, (a)~$k=1.8$, (b)~$k=2$, (c)~$k=3.1$.}
\label{fig2}
\end{figure}

\subsection{A mechanical bifurcation constrains inversion in \emph{P.~californica}}
Based on this model, we expect the inversion mechanics of \emph{P.~californica} to feature a mechanical bifurcation~\figref{fig1}{d}: If the intrinsic curvature in the bend region is sufficiently negative, $k>k_\ast$ say, then the rim of the cell sheet will flip over and it inverts as the wave of cell shape changes propagates to the posterior pole. If $k<k_\ast$ however, then the rim does not flip over, and the cell sheet remains uninverted as the cell shape changes propagate along it, although the rim remains slightly opened because of the moment applied by the negative intrinsic curvature there~\figref{fig1}{d}. A similar bifurcation has been implicated in Ref.~\cite{haas2018embryonic} in the failure of type-A inversion under contractility perturbations and in mutants of \emph{V.~carteri}~\cite{nishii99,nishii03}. There, the lips at anterior pole of the cell sheet~\figrefi{fig1}{a} facilitate flipping over its rim.

We confirm this expectation by computing bifurcation diagrams in $(B,\mathcal{E})$ space for different values of $k$ using the continuation package \texttt{auto-07p}~\citep{auto}, as described in detail in \citeappendix{appB}. At low values of $k$, the cell sheet does not invert on the branch of lowest energy that is connected to the initial, uninverted configuration of the cell sheet~\figref{fig2}{a}. Inverted configurations of lower energy do exist in this bifurcation diagram, but are not connected to the uninverted configuration~\figref{fig2}{a}. At larger values of $k$, this picture changes~\figref{fig2}{b}: The cell sheet now inverts on the branch of lowest energy, while the branch on which the edge of the cell sheet does not flip over has disconnected from the uninverted configuration. These regimes are separated by a bifurcation at $k=k_1$, where, as described in \citeappendix{appB}, a pair of fold points emerges in the bifurcation diagram.

There is in fact a second, relevant bifurcation in this model: At $k=k_2>k_1$, another pair of fold points merges at a branch point on the branch of lowest energy (\citeappendix{appB}) so that inversion is continuous for $k>k_2$ \figref{fig2}{c}, while there is a snap-through at a fold point for $k_1<k<k_2$ as $W$ decreases [represented by the arrow in \textfigref{fig2}{b}]. It is the observation of merging fold points that enables numerical computation of $k_2$ in \texttt{auto-07p} (\citeappendix{appB}).

We thus compute the critical curvatures $k_1,k_2$ for different values of the remaining model parameters $h,L$. Thence we identify, in \textwholefigref{fig3}, the region in $(h,L,k)$ parameter space in which inversion is possible or continuous. It is bounded by the surfaces ${k=k_1(h,L)}$ or ${k=k_2(h,L)}$ and ${k=8/3h}$, of which the latter corresponds to the limit of perfect cell constriction (\citeappendix{appA}), and thus bounds the geometrically inaccessible region of parameter space in which the intrinsic cell shapes would self-intersect. Below this boundary sits the region of strong constriction in which the elastic shell theory underlying model is not formally valid (\citeappendix{appA}).

\begin{figure}[t!]
\includegraphics[width=8.5cm]{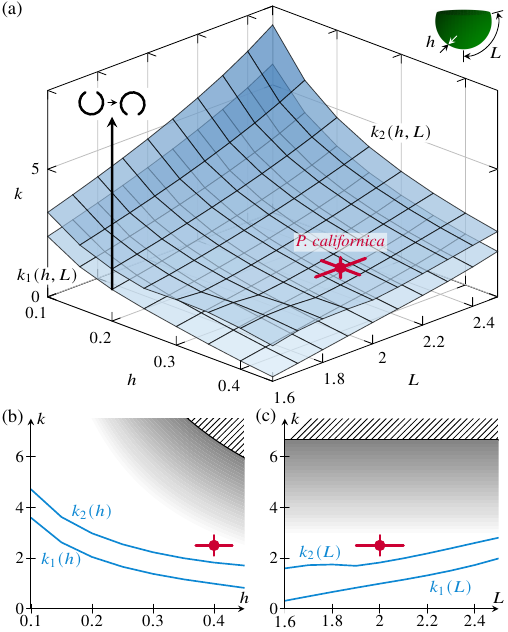}
\centering
\caption{Quantitative model of inversion in \emph{P.~californica}. (a)~Plot, in $(h,L,k)$ space, of the bifurcation boundaries $k_1(h,L),k_2(h,L)$. Inversion is possible (arrow) for $k>k_1(h,L)$ and continuous for ${k>k_2(h,L)}$. Inset: definition of $h,L$. Marker and error bars: parameter values for \emph{P.~californica} estimated in Table~\ref{table1}, falling within the region in which the model allows continuous inversion. (b)~Cross-section through this plot at $L=2.0$ (Table~\ref{table1}). The hatched area ${k>8/(3h)}$ is forbidden geometrically because the intrinsically deformed cell shapes would self-intersect (\citeappendix{appA}). The shading indicates the region of strong cell constriction (\citeappendix{appA}). (c)~Analogous cross section at $h=0.4$ (Table~\ref{table1}).}
\label{fig3}
\end{figure}

\begin{table}[!b]
\caption{Quantitative model of inversion in \emph{P.~californica}. Parameter values estimated from Ref.~\cite{hohn2016distinct}.\label{table1}}
\begin{ruledtabular}
\begin{tabular}{cccc}
parameter & description & estimate\footnote{Details of the measurements and error estimates are given in \citeappendix{appC}.} & source\footnote{Figure(s) of Ref.~\protect{\cite{hohn2016distinct}} on which the estimate of the parameter value is based.}\\
\hline
$L$&extent of cell sheet&$2.0\pm0.1$&Fig. 5(a)\\
$h$&cell sheet thickness&$0.40\pm0.03$&Fig. 5(a)\\
$k$&intrinsic curvature&$2.5\pm0.4$&Figs. 5(b)--5(e)\\
\end{tabular}
\end{ruledtabular}
\end{table}

\subsection{Quantitative model of inversion in \emph{P.~californica}}
To make this model quantitative, we used experimental images of Ref.~\cite{hohn2016distinct} to estimate the values of its dimensionless parameters $L,h,k$ (Table~\ref{table1} and \citeappendix{appC}). Reassuringly, these estimates fall within the region of parameter space in which our model predicts that the cell sheet can invert~\wholefigref{fig3}.

Interestingly, these estimates suggest that inversion happens well above the mechanical limit set by the bifurcation boundary $k=k_1(h,L)$, but rather close to the boundary for continuous inversion set by $k_2(h,L)$. The possibility of a snap-through in inversion has previously been suggested for \emph{V.~carteri}~\cite{viamontes79}.

We also note that the elongated cells of \emph{P.~californica}~\cite{hohn2016distinct} are markedly less wedge-shaped or pointy than the cells that play an analogous role in bending the cell sheet in \emph{Volvox}, i.e., the flask-shaped cells in \emph{V.~carteri}~\citep{viamontes1977cell} or the paddle-shaped cells in \emph{V.~globator}~\citep{hohn11}. This is a first hint that these more drastic cell shape changes, reminiscent of apical constriction in animal development~\cite{sawyer10,keller11}, evolved because of a mechanical requirement to impart higher intrinsic curvatures to the cell sheet.

\begin{figure}[t]
\centering\includegraphics{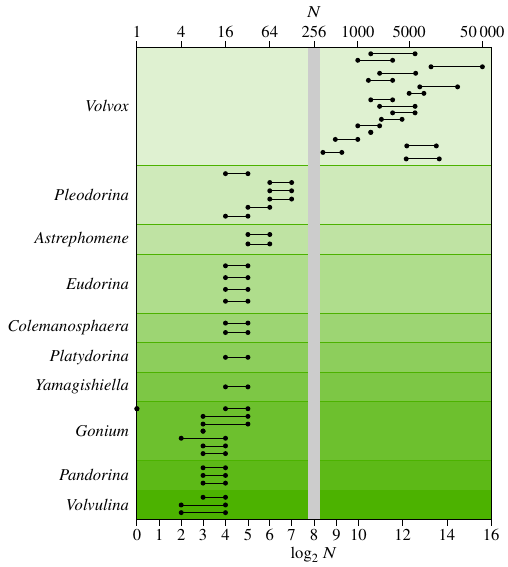}
\caption{Plot of the ranges reported for the numbers of cells of individuals for different species in the families Volvocaceae and Goniaceae, grouped according to genus, based on the Algaebase database~\cite{algaebase} and data from Refs.~\cite{grove1915pleodorina,chodat1931quelques,pocock1955studies,starr1962new,berg1970structure,starr1970volvox,karn1972sexual,akiyama1977illustrations,marchant1977colonymain,ireland1981inversion,nozaki1982morph,nozaki1983morphology,batko1989gonium,nozaki1989morphological,nozaki1989pleodorina,nozaki1990volvulina,nozaki1991pandorina,nozaki1992ultrastructure,nozaki1993asexual,nozaki2001morphology,hallmann2006morphogenesis,nozakimorphology,yamada2008taxonomic,solari2008volvox,hayama2010morphology,iida2011embryogenesis,hohn11,nozaki2011new,isaka2012description,iida2013cleavage,nozaki2014new,nozaki2015morphology,hohn2016distinct,nozaki2017rediscovery,nozaki2018morphology,kimbara2019morphological,nozaki2019morphology,nozaki2019morphologyb,nozaki2022morphology} collected in \citeappendix{appD}. Each line corresponds to one species. The grey bar highlights the lack of species with individuals of 256 cells.}
\label{fig4}
\end{figure}

\subsection{A gap in the family Volvocaceae}
Why then do the larger species of \emph{Volvox} use a different inversion strategy from \emph{Pleodorina}? To address this question, we used the Algaebase database~\cite{algaebase} to survey data on different species of Volvocaceae and the closely related family Goniaceae, reporting in particular the numbers $N$ of cells of individuals of these species. This revealed a curious gap: no individuals with $N=256$ cells have been reported \wholefigref{fig4}. This gap was all the more striking because it separates the ``small'' genera of Volvocaceae from the larger species of \emph{Volvox} in which the numbers of cells are no longer powers of $2$ in general, because of asymmetric cell divisions of the gonidia~\cite{green1981cleavage,kirk2001}.

This suggested an appealing mechanical hypothesis: Could this gap be a consequence of the mechanical bifurcation that we have just described? Could it be that the inversion programme of \emph{Pleodorina} is possible for individuals with 64 or 128 cells, but not for individuals with 256 or 512 cells? This would suggest in turn that the species of \emph{Volvox} had to evolve their more complex inversion programmes to eschew this bifurcation.

\begin{figure}[t]
\centering\includegraphics{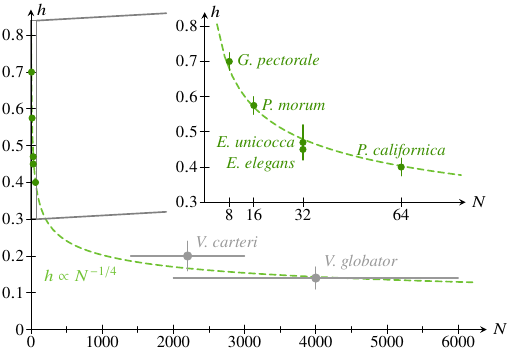}
\caption{Allometry of Volvocaceae and Goniaceae. Illustration of the empirical relation $H\propto N^{-1/4}$ (dashed line) between the non-dimensional cell sheet thickness $h$ and the number $N$ of cells, fitted from micrographs \mbox{\cite{hallmann2006morphogenesis,hohn2016distinct,marchant1977colonymain}} of \emph{Gonium pectorale}, \emph{Pandorina morum}, \emph{Eudorina unicocca}, \emph{Eudorina elegans}, \emph{Pleodorina californica} ($N=\text{8--64}$), as shown in the inset and described in \citeappendix{appC}. The allometric relation is also consistent with estimates from micrographs~\cite{nozaki2018morphology,hohn11} of \emph{Volvox carteri} and \emph{Volvox globator} ($N=\text{1400--6000}$).}
\label{fig5}
\end{figure}

\subsection{Allometry of Volvocaceae}
To address this hypothesis, we needed to be able to extrapolate the model parameters for $N=64$ cells (Table~\ref{table1}) to individuals with $N=128,256,512,1024$ cells. This was made possible by discovering an allometric law \wholefigref{fig5} relating the dimensionless thickness $h$ of the cell sheet to $N$, 
\begin{align}
h=\xi N^{-1/4},\quad\text{where }\xi\approx 1.14\pm 0.06,\label{eq:allom}
\end{align}
in which the prefactor is fitted (\citeappendix{appC}) from estimates obtained from published micrographs for \emph{P.~californica} ($N=64$ cells) \cite{hohn2016distinct}, \emph{Eudorina elegans} ($N=32$) \cite{marchant1977colonymain}, \emph{Eudorina unicocca} ($N=32$) \cite{hallmann2006morphogenesis}, \emph{Pandorina morum} ($N=16$) \cite{hallmann2006morphogenesis}, and \emph{Gonium pectorale} ($N=8$) \cite{hallmann2006morphogenesis}. Strikingly, estimates for \emph{V.~carteri} (\mbox{$N=1400$--$3000$}) \cite{nozaki2018morphology} and \emph{V.~globator} (\mbox{$N=2000$--$6000$}) \cite{hohn11} are also consistent with this allometric law~\wholefigref{fig5}, indicating that it can indeed be used for extrapolation.

\subsection{Extrapolation of cell sheet geometries}
We were left to extrapolate the dimensionless extent of the cell sheet $L$ to species of more than $64$ cells. This extent is set by the packing of the cells onto the spherical cap during the repeated rounds of cell division that the cell sheet undergoes before inversion~\cite{hallmann2006morphogenesis,green1981cleavage}. This packing in turn is constrained by the network of cytoplasmic bridges~\cite{green1981cleavage,green81}. The resulting cleavage patterns have been described in detail in different species of \emph{Volvox} up to the 64-cell stage~\cite{sumper79,green1981cleavage,gilles82,ransick91,wujek06}.

A physics slant on these problems of cell packings has only emerged recently~\cite{imranalsous18,yanni20,day22,fabreges24} through work highlighting their relation to robustness of morphogenesis and the evolution of multicellularity. In particular, cell packings in the early \emph{Drosophila} embryo~\cite{imranalsous18} are constrained by cytoplasmic bridges, too.

\begin{figure}[t]
\centering\includegraphics{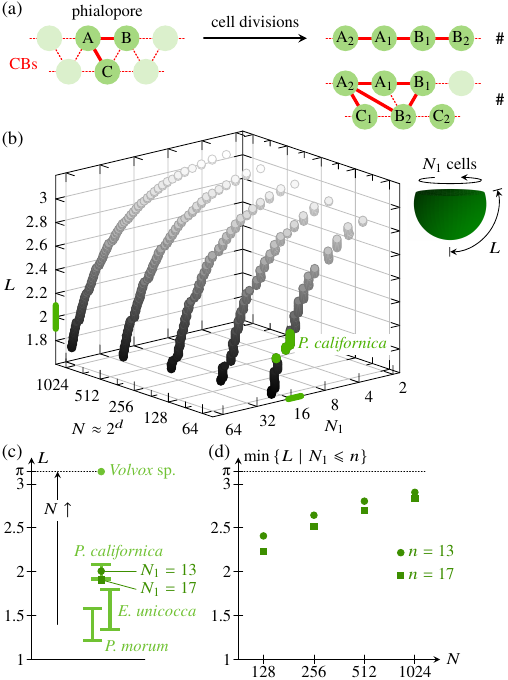}
\caption{Extrapolation of cell sheet geometries. (a) Combinatorial constraints on the arrangement of cells at the phialopore: three cells $\mathrm{A},\mathrm{B},\mathrm{C}$ are connected by cytoplasmic bridges (CBs, red lines), and divide into $\mathrm{A}_1,\mathrm{A}_2,\mathrm{B}_1,\mathrm{B}_2,\mathrm{C}_1,\mathrm{C}_2$, respectively. At most two of $\mathrm{A}_1,\mathrm{A}_2$ and $\mathrm{B}_1,\mathrm{B}_2$ can remain at the phialopore lest a contradiction (marked~\#) arise. This combinatorial argument requires the solid CBs, but not the dotted ones. See text for further explanation. (b)~Plot of the number $N_1$ of cells surrounding the phialopore for all possible packings of ${N\approx 2^d}$ cells into a spherical cap of size $L$. Inset: definitions of $N_1$ and $L$. Packings falling into the range of $L$ estimated in Table~\ref{table1} for \emph{P.~californica} ($N=64$) and the corresponding range of $N_1$ are highlighted. See text for further explanation. (c)~Plot of estimates (Appendix~\ref{appC}) of $L$ from micrographs~\cite{hohn2016distinct,hallmann2006morphogenesis} for \emph{P. californica} (Table~\ref{table1}), \emph{E. unicocca}, and \emph{P. morum}, suggesting that $L$ increases with~$N$ towards the spherical cell sheet of \emph{Volvox} ($L\approx\pi$). Packings with $N=64$ and with $N_1=13$ or $N_1=17$ cells at the phialopore are in the middle and at the lower end of the range of $L$ for \emph{P.~californica} (Table~\ref{table1}). (d)~Minimal values of $L$ for extrapolations of the packings with $N=64$ and $N_1=13$ or $N_1=17$ highlighted in panel~(c) to $N=128,256,512,1024$ cells.}
\label{fig6}
\end{figure}

\subsubsection{Combinatorial constraints on cell packings}
Observations of the cytoplasmic bridge system linking neighbouring cells of \emph{Volvox} during the cleavage phase~\cite{green1981cleavage,green81} suggest the following rules: (1) if cell $\mathrm{A}$ divides into $\mathrm{A}_1,\mathrm{A}_2$, then $\mathrm{A}_1,\mathrm{A}_2$ are connected by cytoplasmic bridges (CBs); (2) if cells $\mathrm{A},\mathrm{B}$ are connected by CBs and divide into  $\mathrm{A}_1,\mathrm{A}_2,\mathrm{B}_1,\mathrm{B}_2$, respectively, then each of $\mathrm{A}_1,\mathrm{A}_2$ is connected by CBs to at least one of $\mathrm{B}_1,\mathrm{B}_2$; (3) CBs connecting different pairs of cells to not cross.

We now focus on the cell packing close to the rim or opening of the cell sheet, the phialopore, assuming that, as is the case for, e.g., \emph{P.~californica}~\cite{hohn2016distinct} or \emph{V.~globator}~\cite{hohn11}, that the cell sheet does not have lips; the following argument does not therefore apply to, e.g., \emph{V.~carteri}~\cite{viamontes1977cell,hallmann2006morphogenesis}.

Consider three cells $\mathrm{A},\mathrm{B},\mathrm{C}$, as shown in~\textfigref{fig6}{a}: cells $\mathrm{A},\mathrm{B}$ are at the phialopore, connected by CBs. Additionally, $\mathrm{A}$ is connected to cell $\mathrm{C}$, which is not at the phialopore. In a round of cell divisions, these cells divide into $\mathrm{A}_1,\mathrm{A}_2,\mathrm{B}_1,\mathrm{B}_2,\mathrm{C}_1,\mathrm{C}_2$, respectively. We claim that at most two of $\mathrm{A}_1,\mathrm{A}_2,\mathrm{B}_1,\mathrm{B}_2$ can remain at the phialopore. 

Indeed, first suppose to the contrary that $\mathrm{A}_1,\mathrm{A}_2,\mathrm{B}_1,\mathrm{B}_2$ are all at the phialopore~\figref{fig6}{a}. By rule (1), $\mathrm{A}_1,\mathrm{A}_2$ and $\mathrm{B}_1,\mathrm{B}_2$ are connected by CBs, while, by rule (2) and without loss of generality, $\mathrm{A}_1$ is connected by CBs to $\mathrm{B}_1$. Hence, again without loss of generality, $\mathrm{A}_2,\mathrm{A}_1,\mathrm{B}_1,\mathrm{B}_2$ are in this order along the phialopore. Thus $\mathrm{A}_2$ neighbours neither $\mathrm{B}_1$ nor $\mathrm{B}_2$ (unless there are no other cells at the phialopore, $N_1=4$, which is not germane to our argument). This contradicts rule (2) and shows that at least one of $\mathrm{A}_1,\mathrm{A}_2,\mathrm{B}_1,\mathrm{B}_2$ is not at the phialopore.

Thus suppose, again without loss of generality, that $\mathrm{A}_2,\mathrm{A}_1,\mathrm{B}_1$ are connected by CBs at the phialopore in this order, but that $\mathrm{B}_2$ is not~\figref{fig6}{a}. By rule (1), $\mathrm{B}_1,\mathrm{B}_2$ are connected by CBs. Moreover, $\mathrm{A}_2,\mathrm{B}_1$ are not neighbours, so, unless $N_1=3$ (which is irrelevant to our argument), $\mathrm{A}_2,\mathrm{B}_2$ are connected by CBs by rule (2). However, rule (2) implies that $\mathrm{A}_1$ must be connected to one of $\mathrm{C}_1,\mathrm{C}_2$, which finally contradicts rule (3). [Indeed, since $\mathrm{A}_1$ is connected to by CBs to $\mathrm{A}_2,\mathrm{B_1}$ by assumption, this contradiction cannot be resolved by $\mathrm{C}_1$ or $\mathrm{C}_2$ being at the phialopore.]

In this way, we have shown that at most two of $\mathrm{A}_1,\mathrm{A}_2,\mathrm{B}_1,\mathrm{B}_2$ can remain at the phialopore. (We note that this argument does not require cells $\mathrm{B},\mathrm{C}$ to be connected by CBs.) The phialopore thus leads to a combinatorial constraint on the cell packings: as the number of cells increases through repeated rounds of cell division, the number of cells at the phialopore cannot increase.

\subsubsection{Analysis of cell packings}
We generate regular packings of $N\approx 2^d$ equally-sized cells into a spherical cap of size $L$ by extending the algorithm of Ref.~\cite{deserno}, and denote by $N_1$ the number of cells surrounding the phialopore in each of these packings. We plot these packings in $(N,N_1,L)$ space in \textfigref{fig6}{b} for $d=\text{6--10}$. 

In particular, \textfigref{fig6}{b} classifies those packings consistent with the range of $L$ estimated in Table~\ref{table1} for \emph{P.~californica} ($d=6$). Packings with $N_1=13$, consistent with Fig.~5(m) of Ref.~\cite{hohn2016distinct}, fall right into this range~\figref{fig6}{c}, although its lower end is also consistent with packings with $N_1=17$. The combinatorial constraint proved above, that~$N_1$ cannot increase if the sheet were to undergo further rounds of cell division, now translates to lower bounds on $L$ for these packings~\figref{fig6}{d}. We note that these lower bounds increase with $N$.

This shows that, if 64-celled \emph{P.~californica} were to undergo further rounds of cell division, then $L$ would increase. Thus the phialopore would shrink compared to the size of the cell sheet and its shape would approximate the spherical one of \emph{Volvox}. This increase of $L$ allowed us to analyse the possibility of inversion in such hypothetical individuals or species below. This increase of $L$ with $N$ is also consistent with that observed across the small genera of Volvocaceae~\figref{fig6}{c}. 

We stress that this argument does not need to, and does not in fact reproduce the cleavage pattern of \emph{Volvox}, for which $L\approx\pi$ throughout the cell division stage~\cite{sumper79,green1981cleavage,wujek06}. This cleavage pattern is not covered by our combinatorial argument, because the cell shapes during early cleavage stages in \emph{Volvox} are elongated~\cite{sumper79,green1981cleavage,wujek06}. The size difference between somatic and germ cells and the lips near the phialopore that occur in some, but not all species of~\emph{Volvox}~\cite{hallmann2006morphogenesis} could also break this argument. However, already by the 64-cell stage, the cell packing of \emph{Volvox}~\cite{sumper79,green1981cleavage,wujek06} does approximate the regular packings of \textfigref{fig6}{b}, as does the cell packing of the smaller genera of Volvocaceae~\cite{hallmann2006morphogenesis} and \emph{P.~californica}~\cite{hohn2016distinct} in particular. This justifies the use of this model to extrapolate our measurements to cell packings resulting from 64-celled \emph{P.~californica} undergoing additional rounds of cleavage.

\subsection{Mechanical constraints on cell sheet folding in Volvocaceae}
We could now extrapolate our measurements for \emph{P.~californica} to hypothetical cell sheets resulting from additional rounds of cell cleavage, assuming that the cell shape changes remain the same (i.e., that the cells remain similarly wedge-shaped). This is to assume that the intrinsic curvature $k_N$ imparted to a cell sheet of $N$ cells by the wedge-shaped cells scales with the cell sheet thickness $h_N$, i.e., $k_{N'}h_{N'}=k_Nh_N$ for any $N,N'$. We plot these extrapolated values, combining the values in Table~\ref{table1}, the allometric scaling of Eq.~\eqref{eq:allom}, and the extrapolated extent of the cell sheet in \textfigref{fig6}{c} in \textwholefigref{fig7}. Since we expect our measurement of $k$, based on the most wedge-shaped cells (\citeappendix{appC}), to be an overestimate, we plot extrapolations of both the mean estimate in Table~\ref{table1} and the lower bound of the corresponding uncertainty interval.

\begin{figure}[b]
\centering\includegraphics{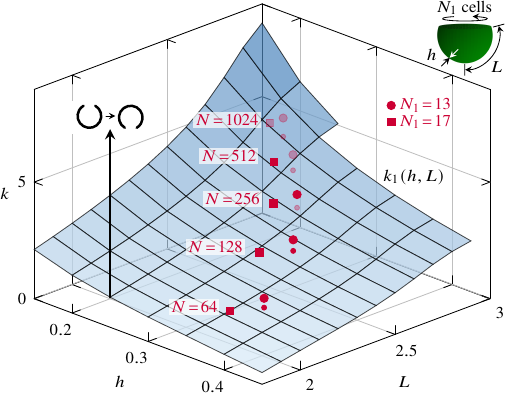}
\caption{Mechanical constraints on cell sheet folding in Volvocaceae. Plot, in $(h,L,k)$ space, of the surface $k=k_1(h,L)$ and of the extrapolation of the parameter measurements for \emph{P.~californica} ($N=64$ cells) to $N=128,256,512,1024$ cells. Disk (square) markers: extrapolation of $L$ for $N_1=13$ ($N_1=17$) from \textfigref{fig6}{d}. Large (small) disk markers: extrapolation of $k$ based on the mean (lower) estimate in Table~\ref{table1}. Inversion fails for sufficiently large $N$, when $k<k_1(h,L)$.}
\label{fig7}
\end{figure}

Reassuringly, all estimates for $N=128$ cells lie above the surface $k=k_1(h,L)$, i.e., within the region in which the model allows inversion, consistently with descriptions of inverting 128-celled individuals of different species in the genus \emph{Pleodorina}~\cite{hohn2016distinct,nozaki1989pleodorina,nozaki2017rediscovery}. Conversely, by the 256-cell stage, the first estimate falls below $k=k_1(h,L)$, and all estimates lie below this surface for $N=1024$.

Our model thus predicts that the bifurcation at $k=k_1(h,L)$ sets $d=8$--$10$ ($256$--$1024$ cells) as the mechanical upper limit of the inversion programme of \emph{P.~californica}. In other words, this programme would fail if it were to undergo more rounds of cell division. This agrees with the gap in the Volvocaceae at $d=8$ and with alternative inversion programmes appearing in the genus \emph{Volvox} at $d=9$--$10$ \wholefigref{fig4}. In turn, this hints that the different programmes of inversion in \emph{Volvox} \figrefi{fig1}{a} and, in particular, the higher degree of cell constriction in its bend region mentioned above arose from an evolutionary requirement to circumvent this bifurcation.

\section{\uppercase{Discussion and conclusion}}
In this paper, we have discovered that a mechanical bifurcation constrained the evolution of cell sheet folding in the family Volvocaceae: the simple programme of inversion in the genus \emph{Pleodorina}~\figref{fig1}{b} runs into a mechanical bifurcation; the more complex inversion programmes in the genus \emph{Volvox}~\figrefi{fig1}{a} are therefore, the model suggests, evolutionary innovations made necessary by this bifurcation.

This indicates a new role of mechanical bifurcations in development, of constraining the evolution of morphogenetic processes. Future work will need to identify additional examples of this role in different biological systems. This requires, in particular, quantitative mechanical models of these developmental processes. In the minimal model that is the development of Volvocaceae, this was enabled by an elastic model of inversion that has been tested extensively in \emph{Volvox}~\cite{haas15,haas2018embryonic,haas2018noisy,haas25}. Still, it might be argued that the model is limited by the small amount of experimental data underlying the quantification of the model parameters (\citeappendix{appC}). It is therefore fortunate that the underlying bifurcation \wholefigref{fig7} is ``unavoidable'' in parameter space as long as the qualitative behaviour implied by our quantitative estimates continues to hold: the cell sheet becomes (relatively) thinner and more spherical as the number of cells increases.

More physically, our results have highlighted the role of apical constriction in overcoming the mechanical bifurcation through cell shape changes that impart more intrinsic curvature to the cell sheet. Our mechanical model (\citeappendix{appA}) has stressed the role of the mechanics of constriction: in the limit of large intrinsic curvatures associated with strongly constricted cells, the mechanics of the cell sheet change quantitatively and even qualitatively~\cite{haas2021morphoelasticity}. Still, true constriction is a geometric singularity that displays interesting mechanical consequences even in minimal models~\cite{guharay25}. Obtaining a more general continuum theory of cell constriction in tissues therefore remains an interesting avenue for future theoretical work.

\begin{acknowledgments}
We thank Yuan He for preliminary computations at an early stage of this work and Steph Höhn and Hisayoshi Nozaki for discussions. We further thank Steph Höhn for comments on the manuscript and for providing the figure panels from Ref.~\cite{hohn2016distinct} reused in Figs.~\ref{fig1} and~\ref{fig10}. We are grateful for funding from the Max Planck Society. V.~T.~gratefully acknowledges additional support from the Erasmus+ programme of the European Union and from École Normale Supérieure.
\end{acknowledgments}

\section*{Data availability}
\texttt{fortran} code implementing the governing equations for the elastic problem (derived in Appendix~\ref{appA}), code for the numerical calculation of the bifurcations in \texttt{auto-07p}~\cite{auto}, and \textsc{Matlab} (The MathWorks, Inc.) code for generating cell packings based on Ref.~\cite{deserno} are openly available at Ref.~\footnote{Code is available at \href{https://doi.org/10.5281/zenodo.19046776}{\texttt{doi.org/10.5281/zenodo.19046776}}.}.

\appendix
\titleformat{\section}{\normalfont\bfseries\centering\small}{APPENDIX \thesection:}{0.5em}{}
\section{\uppercase{Elastic model}}\label{appA}
We describe the cell sheet of \emph{Pleodorina} as an axisymmetric elastic shell, in agreement with earlier mechanical models of \emph{Volvox}~\cite{hohn2015inversion,haas15,haas2018noisy,haas2018embryonic,haas25}. Cell shape changes endow the elastic shell with intrinsic stretches $f_s^0,\smash{f_\phi^0}$ and intrinsic curvatures $\kappa_s^0,\smash{\kappa_\phi^0}$ in the longitudinal and azimuthal directions, respectively. To represent the large intrinsic curvatures associated with wedge-shaped cells, we use the shell theory for ``large bending deformations'' of Ref.~\cite{haas2021morphoelasticity}. We note that, however, Ref.~\cite{haas2021morphoelasticity} derived a shell theory for $\kappa_s^0\gg\smash{\kappa_\phi^0}$ in its main text,  motivated by the paddle-shaped cells in the bend region of \emph{V. globator}~\cite{hohn11}, while $\kappa_s^0\approx\smash{\kappa_\phi^0}$ are both large in the bend region of \emph{P. californica}~\cite{hohn2016distinct} or \emph{V. carteri}~\cite{viamontes79,haas2018embryonic}. As noted above, the cell size measurements of Ref.~\cite{hohn2016distinct} suggest that the cell shape changes of \emph{P.~californica} do not lead to contraction or expansion within the plane of the cell sheet. For this reason, we take $f_s^0=\smash{f_\phi^0}=1$. An axisymmetric shell theory valid in this limit can in principle be obtained from the theory for general curved surfaces in Appendix~A of Ref.~\cite{haas2021morphoelasticity}, but it is more convenient to derive the axisymmetric shell theory from first principles.

\begin{figure}[b]
\centering\includegraphics{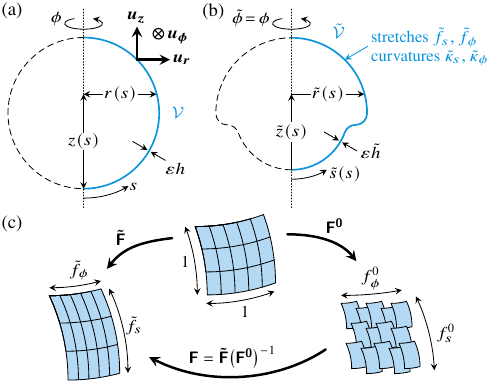}
\caption{Morphoelastic shell theory. (a) Undeformed configuration $\mathcal{V}$ of an axisymmetric shell of thickness $\varepsilon h$, where $\varepsilon\ll 1$. Its midsurface is determined by the coordinates $r(s),z(s)$, where $s$ is arclength of the cross-section, with respect to the basis $\{\vec{u_r},\vec{u_\phi},\vec{u_z}\}$ of cylindrical polars. (b) Configuration $\smash{\tilde{\mathcal{V}}}$ of the shell after a torsionless deformation: the deformed shell has thickness $\varepsilon\tilde{h}$ and arclength $\tilde{s}(s)$. The coordinates of its midsurface with respect to cylindrical polars are $\tilde{r}(s),\tilde{z}(s)$, defining the stretches $\tilde{f}_s,\tilde{f}_\phi$ and curvatures $\tilde{\kappa}_s,\tilde{\kappa}_\phi$. (c)~The deformation from $\mathcal{V}$ (top) to $\smash{\tilde{\mathcal{V}}}$ (left), with deformation gradient $\smash{\btilde{\tens{F}}}$, decomposes into an intrinsic deformation, with deformation gradient $\smash{\tens{F^0}}$ and an elastic deformation, with deformation gradient $\smash{\tens{F}=\btilde{\tens{F}}\bigl(\tens{F^0}\bigr)^{-1}}$. The first defines the intrinsic configuration $\mathcal{V}^0$ (right), with intrinsic stretches $\smash{f_s^0,f_{\smash{\phi}}^0}$ and intrinsic curvatures $\smash{\kappa_s^0,\kappa_{\smash{\phi}}^0}$. See text for further explanation. Figure panels redrawn from Ref.~\cite{haas2021morphoelasticity}.}
\label{fig8}
\end{figure}

\subsection{Derivation of the elastic shell theory}
We thus consider the axisymmetric, torsionless deformations of an axisymmetric elastic shell of small thickness $\varepsilon h$, which enables asymptotic expansion in $\varepsilon\ll 1$. 
\subsubsection{Geometry of axisymmetric shells}
The axisymmetric undeformed configuration $\mathcal{V}$ of the shell~\figref{fig8}{a} is defined by the cross section of its midsurface, with arclength $s$ and coordinates $r(s),z(s)$ with respect to cylindrical polars. The tangent angle to the undeformed midsurface is $\psi(s)$, where 
\begin{align}
&r'(s)=\cos{\psi(s)},&&z'(s)=\sin{\psi(s)},
\end{align}
in which the dash denotes differentiation with respect to $s$. The curvatures of the undeformed midsurface are thus
\begin{align}
&\varkappa_s(s)=\psi'(s),&&\varkappa_\phi(s)=\dfrac{\sin{\psi}(s)}{r(s)}.
\end{align}
Correspondingly, the cross section of the midsurface of the deformation configuration $\tilde{\mathcal{V}}$~\figref{fig8}{b} has arclength $\tilde{s}(s)$, and coordinates $\tilde{r}(s),\tilde{z}(s)$. These define the stretches~\cite{haas2021morphoelasticity}
\begin{align}
&\tilde{f}_s(s)=\tilde{s}'(s),&&\tilde{f}_\phi(s)=\dfrac{\tilde{r}(s)}{r(s)}.
\end{align}
With this, we define the tangent angle $\tilde{\psi}(s)$ of the deformed configuration by
\begin{align}
&\tilde{r}'(s)=\tilde{f}_s\cos{\tilde{\psi}(s)},&& \tilde{z}'(s)=\tilde{f}_s\sin{\tilde{\psi}(s)},\label{eq:drdz}
\end{align}
The curvatures of the deformed shell are thus~\cite{haas2021morphoelasticity}
\begin{align}
&\tilde{\kappa}_s(s)=\dfrac{\tilde{\psi}'(s)}{\tilde{f}_s(s)},&& \tilde{\kappa}_\phi(s)=\dfrac{\sin{\tilde{\psi}(s)}}{\tilde{r}(s)}.\label{eq:dpsi}
\end{align}
The deformation of $\mathcal{V}$ to $\tilde{\mathcal{V}}$ is described by a deformation gradient $\btilde{\tens{F}}$. By the fundamental relation of morphoelasticity~\cite{goriely2017mathematics,ambrosi19}, this decomposes multiplicatively~\figref{fig8}{c} into an intrinsic deformation gradient $\tens{F^0}$ and an elastic deformation gradient $\tens{F}=\btilde{\tens{F}}\smash{\bigl(\tens{F^0}\bigr)^{-1}}$. The former introduces the intrinsic configuration $\mathcal{V}^0$ of the shell~\figref{fig8}{c}. This has thickness $\varepsilon h^0$, and its midsurface is associated with intrinsic stretches $f_s^0,f_\phi^0$, and intrinsic curvatures $\kappa_s^0,\kappa_\phi^0$.

As explained in detail in Ref.~\cite{haas2021morphoelasticity}, we can complete the description of $\mathcal{V}^0$ by introducing the intrinsic coordinate $\zeta^0$ transverse to the midsurface $\zeta^0=0$, such that ${-h^0/2\leq \zeta^0\leq h^0/2}$. The volume element of the intrinsic configuration is then~\cite{haas2021morphoelasticity}
\begin{align}
\mathrm{d}V^0=\varepsilon f_s^0f_\phi^0\bigl(1-\kappa_s^0\zeta^0\bigr)\bigl(1-\kappa_\phi^0\zeta^0\bigr)\,r\,\mathrm{d}s\,\mathrm{d}\phi\,\mathrm{d}\zeta^0.\label{eq:dV0}
\end{align}
To the transverse coordinate $\zeta^0$ of $\mathcal{V}^0$ corresponds the transverse coordinate $\zeta$ of $\mathcal{V}$, with $\zeta^0=0$ mapping to $\zeta=0$, so that $-h^-\leq \zeta\leq h^+$, where $h^-+h^+=h$. Similarly, the transverse coordinate in $\tilde{\mathcal{V}}$ is $\tilde{\zeta}$, with $\zeta^0=0$  corresponding to $\tilde{\zeta}=0$. Moreover, as noted in Ref.~\cite{haas2021morphoelasticity}, points off the midsurfaces may be displaced parallel to the corresponding points in $\mathcal{V}$. We denote this displacement by $\varepsilon\tilde{\varsigma}$, so that $\tilde{\varsigma}=0$ on $\zeta^0=0$.

The deformation gradients can be expressed in terms of the geometric quantities introduced above. Applying the chain rule to write Eqs.~(23)--(25) of Ref.~\cite{haas2021morphoelasticity} in terms of $\smash{\zeta\bigl(s,\zeta^0\bigr)}$, $\smash{\tilde{\zeta}\bigl(s,\zeta^0\bigr)}$, $\smash{\tilde{\varsigma}\bigl(s,\zeta^0\bigr)}$ and using commata to denote differentiation, we obtain
\begin{widetext}
\vspace{-16pt}
\begin{align}
\btilde{\tens{F}}&=\left(\begin{array}{ccc}
\dfrac{\tilde{f}_s\left(1-\varepsilon\tilde{\kappa}_s \tilde{\zeta}\right)+\varepsilon \tilde{\varsigma}_{,s}}{1-\varepsilon\varkappa_s\zeta}&0&\dfrac{\tilde{\varsigma}_{,\zeta^0}}{\zeta_{,\zeta^0}}\\
0&\dfrac{\tilde{f}_\phi\left(1-\varepsilon\tilde{\kappa}_\phi \tilde{\zeta}\right)+\dfrac{\varepsilon \tilde{\varsigma}}{r}\cos{\tilde{\psi}}}{1-\varepsilon\varkappa_\phi\zeta}&0\\
\dfrac{\varepsilon\left(\tilde{\zeta}_{,s}+\tilde{f}_s\tilde{\kappa}_s \tilde{\varsigma}\right)}{1-\varepsilon\varkappa_s\zeta}&0&\dfrac{\tilde{\zeta}_{,\zeta^0}}{\zeta_{,\zeta^0}}
\end{array}\right),&&\tens{F^0}=\left(\begin{array}{ccc}
\dfrac{f_s^0\left(1-\varepsilon\kappa_s^0\zeta^0\right)}{1-\varepsilon\varkappa_s\zeta}&0&0\\
0&\dfrac{f_\phi^0\bigl(1-\varepsilon\kappa_\phi^0\zeta^0\bigr)}{1-\varepsilon\varkappa_\phi\zeta}&0\\
\dfrac{\varepsilon \zeta_{,s}}{(1-\varepsilon\varkappa_s\zeta)\zeta_{,\zeta^0}}&0&\dfrac{1}{\zeta_{,\zeta^0}}
\end{array}\right),\label{eq:FF0}
\end{align}
whence
\begin{align}
\tens{F}=\btilde{\tens{F}}\bigl(\tens{F^0}\bigr)^{-1}=\left(\begin{array}{ccc}
\dfrac{\tilde{f}_s\left(1-\varepsilon\tilde{\kappa}_s \tilde{\zeta}\right)+\varepsilon\left(\tilde{\varsigma}_{,s}+\dfrac{\tilde{\varsigma}_{,\zeta^0}\zeta_{,s}}{\zeta_{,\zeta^0}}\right)}{f_s^0\left(1-\varepsilon\kappa_s^0\zeta^0\right)}&0&\tilde{\varsigma}_{,\zeta^0}\\
0&\dfrac{\tilde{f}_\phi\left(1-\varepsilon\tilde{\kappa}_\phi \tilde{\zeta}\right)+\dfrac{\varepsilon\tilde{\varsigma}}{r}\cos{\tilde{\psi}}}{f_\phi^0\bigl(1-\varepsilon\kappa_\phi^0\zeta^0\bigr)}&0\\
\dfrac{\varepsilon\left(\tilde{\zeta}_{,s}+\tilde{f}_s\tilde{\kappa}_s\tilde{\varsigma}-\dfrac{\zeta_{,s}\tilde{\zeta}_{,\zeta^0}}{\zeta_{,\zeta^0}}\right)}{f_s^0\left(1-\varepsilon\kappa_s^0\zeta^0\right)}&0&\tilde{\zeta}_{,\zeta^0}
\end{array}\right),\label{eq:F}
\end{align}
\vspace{-10pt}
\end{widetext}
\subsubsection{Elasticity of the shell}
The right Cauchy--Green tensor~\cite{goriely2017mathematics} associated to the elastic deformation gradient is $\tens{C}=\tens{F}^\top\tens{F}$, so, with the simplest, neo-Hookean constitutive relations, the elastic energy is
\begin{align}
\mathcal{E}=\iiint_{\mathcal{V}^0}{e\,\mathrm{d}V^0},\quad\text{with }e=\dfrac{C}{2}(\mathcal{I}_1-3), \label{eq:E}
\end{align}
wherein $C>0$ is a material parameter, $\mathcal{I}_1=\operatorname{tr}(\tens{C})$ is the first invariant of $\tens{C}$, and $\mathrm{d}V^0$ is the volume element of the intrinsic configuration $\mathcal{V}^0$ of the shell. We introduce the Piola--Kirchhoff tensor $\smash{\tens{Q}=\tens{F}\bigl(\tens{F^0}\bigr)^{-\top}-p\btilde{\tens{F}}^{-\top}}$, where $p$ is the Lagrange multiplier that imposes the condition $\det{\tens{F}}=1$ of incompressibility. The deformed configuration of the shell then satisfies~\cite{haas2021morphoelasticity}
\begin{align}
\operatorname{Div}{\tens{Q}^\top}=\dfrac{(\tens{Q}\vec{n})_{,\zeta^0}}{\varepsilon\zeta_{,\zeta^0}}+\vec{\nabla}\cdot\tens{Q}^\top=\vec{0},\label{eq:Cauchy}
\end{align}
subject to $\tens{Q}\vec{n}=\vec{0}$ on $\zeta^0=\pm h^0/2$. In Eq.~\eqref{eq:Cauchy}, the divergence operator (with respect to the undeformed configuration of the shell) is split into components parallel and transverse to the undeformed midsurface.
\subsubsection{Asymptotic scalings}
We now impose the scalings of the intrinsic stresses and curvatures discussed above for inversion of \emph{P.~californica}, viz.,
\begin{align}
&f_s^0(s)=f_\phi^0(s)=1,&&\kappa_s^0(s)=\kappa_\phi^0(s)=\dfrac{\kappa^0(s)}{\varepsilon}.
\end{align}
Accordingly, we define the shell strains $E_s,E_\phi$ and the curvature strains $L_s,L_\phi$ by writing
\begin{align}
&\tilde{f}_s(s)=1+\varepsilon E_s(s),&&\tilde{f}_\phi(s)=1+\varepsilon E_\phi(s),
\end{align}
and
\begin{align}
&\tilde{\kappa}_s(s)=\dfrac{\kappa^0(s)}{\varepsilon}+L_s(s),&\tilde{\kappa}_\phi(s)=\dfrac{\kappa^0(s)}{\varepsilon}+L_\phi(s).
\end{align}
\subsubsection{Asymptotic expansion}
We solve the Cauchy equation~\eqref{eq:Cauchy} together with the incompressibility condition $\det{\tens{F}}=1$ by asymptotic expansion in $\varepsilon\ll1$, by positing
\begin{subequations}
\begin{align}
\tilde{\zeta}\bigl(s,\zeta^0\bigr)&=Z_{(0)}\bigl(s,\zeta^0\bigr)+\varepsilon Z_{(1)}\bigl(s,\zeta^0\bigr)+O\bigl(\varepsilon^2\bigr),\\
\tilde{\varsigma}\bigl(s,\zeta^0\bigr)&=S_{(0)}\bigl(s,\zeta^0\bigr)+O(\varepsilon),\\
p\bigl(s,\zeta^0\bigr)&=p_{(0)}\bigl(s,\zeta^0\bigr)+O(\varepsilon).
\end{align}
\end{subequations}
Now writing $\smash{\tens{Q}=\tens{Q_{(0)}}+O(\varepsilon)}$, Eq.~\eqref{eq:Cauchy} yields, at leading order, $(\tens{Q_{(0)}}\vec{n})_{,\zeta^0}=\vec{0}$ subject to $\tens{Q_{(0)}}\vec{n}=\vec{0}$ on $\zeta^0=\pm h^0/2$, so $\tens{Q_{(0)}}\vec{n}\equiv\vec{0}$. Using \textsc{Mathematica} (Wolfram, Inc.), this condition yields
\begin{subequations}\label{eq:lo}
\begin{align}
S_{(0),\zeta^0}&=-\dfrac{p_{(0)}\kappa^0S_{(0)}}{\bigl(1-\kappa^0Z_{(0)}\bigr)Z_{(0),\zeta^0}-\kappa^0S_{(0)}S_{(0),\zeta^0}},\\
Z_{(0),\zeta^0}&=\dfrac{p_{(0)}\bigl(1-\kappa^0Z_{(0)}\bigr)}{\bigl(1-\kappa^0Z_{(0)}\bigr)Z_{(0),\zeta^0}-\kappa^0S_{(0)}S_{(0),\zeta^0}}.
\end{align}
The leading-order expansion of $\det{\tens{F}}=1$ gives
\begin{align}
\bigl(1-\kappa^0Z_{(0)}\bigr)Z_{(0),\zeta^0}-\kappa^0S_{(0)}S_{(0),\zeta^0}=\dfrac{\bigl(1-\kappa^0\zeta^0\bigr)^2}{1-\kappa^0Z_{(0)}}.
\end{align}
\end{subequations}
Equations~\eqref{eq:lo} define a system of algebraic equations for $Z_{(0),\zeta^0},S_{(0),\zeta^0},p_{(0)}$, with solution
\begin{subequations}
\begin{align}
Z_{(0),\zeta^0}&=\dfrac{\bigl(1-\zeta^0\kappa^0\bigr)^2}{\bigl(1-Z_{(0)}\kappa^0\bigr)^2+\bigl(S_{(0)}\kappa^0\bigr)^2},\label{eq:Z0}\\
S_{(0),\zeta^0}&=-\dfrac{S_{(0)}\kappa^0\bigl(1-\zeta^0\kappa^0\bigr)^2}{\bigl(1-Z_{(0)}\kappa^0\bigr)\left[\bigl(1-Z_{(0)}\kappa^0\bigr)^2+\bigl(S_{(0)}\kappa^0\bigr)^2\right]},\label{eq:S0}\\
p_{(0)}&=\dfrac{\bigl(1-\zeta^0\kappa^0\bigr)^4}{\bigl(1-Z_{(0)}\kappa^0\bigr)^2\left[\bigl(1-Z_{(0)}\kappa^0\bigr)^2+\bigl(S_{(0)}\kappa^0\bigr)^2\right]}.\label{eq:p0}
\end{align}
\end{subequations}
From Eqs.~\eqref{eq:Z0} and \eqref{eq:S0},
\begin{subequations}
\begin{align}
\dfrac{S_{(0),\zeta^0}}{S_{(0)}}+\dfrac{\lambda_0Z_{(0),\zeta^0}}{1-Z_{(0)}\kappa^0}=0,
\end{align}
which integrates to
\begin{align}
\dfrac{S_{(0)}}{1-\kappa^0Z_{(0)}}=\text{const.}=0,
\end{align}
\end{subequations}
using $S_{(0)}=Z_{(0)}=0$ on $\zeta^0=0$. Thus $S_{(0)}\equiv 0$, so Eq.~\eqref{eq:Z0} becomes 
\begin{align}
\bigl(1-Z_{(0)}\kappa^0\bigr)^2Z_{(0),\zeta^0}=\bigl(1-\zeta^0\kappa^0\bigr)^2,
\end{align}
which, using $Z_{(0)}=0$ on $\zeta^0=0$ again, integrates to $Z_{(0)}=\zeta^0$. Finally, $p_{(0)}=1$ from Eq.~\eqref{eq:p0}. At order $O(\varepsilon)$, $\det{\tens{F}}=1$ then becomes
\begin{subequations}
\begin{align}
Z_{(1),\zeta^0}-\dfrac{2\kappa^0Z_{(1)}}{1-\kappa^0\zeta^0}+E_s+E_\phi-\dfrac{\zeta^0(L_s+L_\phi)}{1-\kappa^0\zeta^0}=0,
\end{align}
a differential equation for $Z_{(1)}$, the solution of which that satisfies $Z_{(1)}=0$ on $\zeta^0=0$ is
\begin{align}
Z_{(1)}&=\dfrac{3-3\kappa^0\zeta^0+\bigl(\kappa^0\zeta^0\bigr)^2}{3\bigl(1-\kappa^0\zeta^0\bigr)^2}(E_s+E_\phi)\nonumber\\
&\qquad+\dfrac{\bigl(\zeta^0\bigr)^2\bigl(3-2\kappa^0\zeta^0\bigr)}{6\bigl(1-\kappa^0\zeta^0\bigr)^2}(L_s+L_\phi).    
\end{align}    
\end{subequations}
With this, we obtain
\begin{align}
\tens{F}=\left(\begin{array}{ccc}
1+\varepsilon a_{(1)}&0&\varepsilon v_{(1)}\\
0&1+\varepsilon b_{(1)}&0\\
\varepsilon w_{(1)}&0&1+\varepsilon c_{(1)}
\end{array}\right)+O\bigl(\varepsilon^2\bigr),\label{eq:F2}
\end{align}
in which
\begin{subequations}
\begin{align}
a_{(1)}&=E_s-\dfrac{\zeta^0L_s}{1-\kappa^0\zeta^0}-\dfrac{\bigl(\zeta^0\bigr)^2\bigl(3-2\kappa^0\zeta^0\bigr)}{6\bigl(1-\kappa^0\zeta^0\bigr)^3}(L_s+L_\phi)\nonumber\\
&\qquad+\dfrac{\kappa^0\zeta^0\left[3-3\kappa^0\zeta^0+\bigl(\kappa^0\zeta^0\bigr)^2\right]}{3\bigl(1-\kappa^0\zeta^0\bigr)^3}(E_s+E_\phi),\\
b_{(1)}&=E_\phi-\dfrac{\zeta^0L_\phi}{1-\kappa^0\zeta^0}-\dfrac{\bigl(\zeta^0\bigr)^2\bigl(3-2\kappa^0\zeta^0\bigr)}{6\bigl(1-\kappa^0\zeta^0\bigr)^3}(L_s+L_\phi)\nonumber\\
&\qquad+\dfrac{\kappa^0\zeta^0\left[3-3\kappa^0\zeta^0+\bigl(\kappa^0\zeta^0\bigr)^2\right]}{3\bigl(1-\kappa^0\zeta^0\bigr)^3}(E_s+E_\phi),
\end{align}
\end{subequations}
and the values of $c_{(1)},v_{(1)},w_{(1)}$ are of no consequence. As explained in Ref.~\cite{haas2021morphoelasticity}, the formal expansion \eqref{eq:F2} implies 
\begin{align}
\mathcal{I}_1-3=4\varepsilon^2\left[a_{(1)}^2+a_{(1)}b_{(1)}+b_{(1)}^2\right]+O\bigl(\varepsilon^3\bigr).
\end{align}
As discussed in Ref.~\cite{haas2021morphoelasticity}, our shell theory is more naturally expressed in terms of
\begin{subequations}
\begin{align}
K_s&=\tilde{f}_s\tilde{\kappa}_s-\kappa_s^0=L_s+\kappa^0E_s+O(\varepsilon),\\
K_\phi&=\tilde{f}_\phi\tilde{\kappa}_\phi-\kappa_\phi^0=L_\phi+\kappa^0E_\phi+O(\varepsilon),
\end{align}
\end{subequations}
because, unlike $L_s,L_\phi$, these modified curvature strains vanish for ``pure bending'' deformations. Moreover, from Eq.~\eqref{eq:dV0},
\begin{align}
\mathrm{d}V^0=\varepsilon\bigl(1-\kappa^0\zeta^0\bigr)^2\,r\,\mathrm{d}s\,\mathrm{d}\phi\,\mathrm{d}\zeta^0+O\bigl(\varepsilon^2\bigr),
\end{align}
so Eq.~\eqref{eq:E} yields
\begin{align}
\mathcal{E}=2\pi\varepsilon^3\int_{\mathcal{C}}{\hat{e}\,r\,\mathrm{d}s}+O\bigl(\varepsilon^4\bigr),\label{eq:E2}
\end{align}
where the integration is over the curve $\mathcal{C}$ generating the axisymmetric undeformed midsurface, and in which
\begin{align}
\hat{e}=\dfrac{C}{2}\int_{-h^0/2}^{h^0/2}{4\left[a_{(1)}^2+a_{(1)}b_{(1)}+b_{(1)}^2\right]\bigl(1-\kappa^0\zeta^0\bigr)^2\,\mathrm{d}\zeta^0}.\label{eq:eshell}
\end{align}
We obtain a non-asymptotic form of Eq.~\eqref{eq:E2} by setting $\varepsilon=1$ or, more formally, by replacing $\varepsilon h\to h$, $\varepsilon E_s\to E_s$, $\varepsilon E_\phi\to E_\phi$, $\kappa^0/\varepsilon\to\kappa^0$. 
\subsubsection{Intrinsic volume conservation}
To make further progress, we need to determine $h^0$, which follows, as in Ref.~\cite{haas2021morphoelasticity}, by imposing the constraint of intrinsic volume conservation, $\det{\tens{F^0}}=1$. From the second of Eqs.~\eqref{eq:FF0}, at leading order,
\begin{subequations}
\begin{align}
\dfrac{\bigl(1-\kappa^0\zeta^0\bigr)^2}{\zeta_{,\zeta^0}}=1,
\end{align}
which, using $\zeta^0=0\Leftrightarrow\zeta =0$, integrates to
\begin{align}
\zeta^0-\kappa^0\bigl(\zeta^0\bigr)^2+\bigl(\kappa^0\bigr)^2\dfrac{\bigl(\zeta^0\bigr)^3}{3}=\zeta.
\end{align}
\end{subequations}
Now $\zeta^0=\pm h^0/2\Leftrightarrow\zeta=\pm h^\pm$ implies
\begin{subequations}
\begin{align}
h^\pm=\dfrac{h^0}{2}\mp\dfrac{\kappa^0}{4}\bigl(h^0\bigr)^2+\dfrac{\bigl(\kappa^0\bigr)^2}{24}\bigl(h^0\bigr)^3.
\end{align}
By definition, $h=h^++h^-$, so
\begin{align}
h = h^0+\dfrac{\bigl(\kappa^0\bigr)^2}{12}\bigl(h^0\bigr)^3.\label{eq:hh0}
\end{align}
\end{subequations}
It will turn out to be convenient to define
\begin{align}
&\eta=\dfrac{\kappa^0h}{2},&&H=\dfrac{\kappa^0h^0}{2}.
\end{align}
From Eq.~\eqref{eq:hh0}, 
\begin{align}
\eta=H+\dfrac{H^3}{3}.\label{eq:etas}
\end{align}
This cubic equation can be solved in closed form for numerical evaluation of $H$ from $\eta$, and hence from $\kappa^0$ and $h$.
\subsubsection{Elastic shell theory}
With these definitions, we can evaluate the integral in Eq.~\eqref{eq:eshell} in closed form using \textsc{Mathematica}. We find
\newpage
\phantom{.}\vspace{-16pt}
\begin{align}
\hat{e}&=\dfrac{C}{2}\Bigl\{h\bigl(\alpha_{ss}E_s^2+2\alpha_{s\phi}E_sE_\phi+\alpha_{\phi\phi}E_\phi^2\bigr)\nonumber\\
&\hspace{9mm}+2h^2\left[\beta_{ss}E_sK_s+\beta_{s\phi}(E_sK_\phi+E_\phi K_s)+\beta_{\phi\phi}E_\phi K_\phi\right]\nonumber\\
&\hspace{9mm}+h^3\bigl(\gamma_{ss}K_s^2+2\gamma_{s\phi}K_s K_\phi+\gamma_{\phi\phi}K_\phi^2\bigr)\Bigr\},
\end{align}
in which
\begin{subequations}\label{eq:coeffs}
\begin{align}
\alpha_{ss}&=\alpha_{\phi\phi}=\frac{3\bigl(4-2H^2+3H^4-H^6\bigr)}{\left(1-H^2\right)^3\bigl(3+H^2\bigr)},\\
\alpha_{s\phi}&=\frac{3\bigl(2+4H^2-3H^4+H^6\bigr)}{\left(1-H^2\right)^3\bigl(3+H^2\bigr)},\\
\beta_{ss}&=\beta_{\phi\phi}=\frac{9\tanh^{-1}{H}}{2\left[H\bigl(3+H^2\bigr)\right]^2}-\frac{3}{2H\left(1-H^2\right)^3\bigl(3+H^2\bigr)},\\
\beta_{s\phi}&=\dfrac{15\tanh^{-1}{H}}{2\left[H\bigl(3+H^2\bigr)\right]^2}-\dfrac{15-7H^2+15H^4-9H^6+2H^8}{2H\left(1-H^2\right)^3\bigl(3+H^2\bigr)^2},\\
\gamma_{ss}&=\gamma_{\phi\phi}=\dfrac{3\bigl(3+4H^2\bigr)}{4H^2\bigl(3+H^2\bigr)^3}+\dfrac{3}{4H^2\bigl(1-H^2\bigr)^3\bigl(3+H^2\bigr)^2}\nonumber\\
&\qquad\qquad-\dfrac{9\tanh^{-1}{H}}{2\left[H\bigl(3+H^2\bigr)\right]^3},\\
\gamma_{s\phi}&=\dfrac{3\bigl(6-10H^2+15H^4-9H^6+2H^8\bigr)}{4H^2\left[\bigl(1-H^2\bigr)\bigl(3+H^2\bigr)\right]^3}\nonumber\\
&\qquad-\dfrac{9\tanh^{-1}{H}}{2\left[H\bigl(3+H^2\bigr)\right]^3}.
\end{align} 
\end{subequations}
\subsection{Governing equations}
The governing equations for our numerical solution are the Euler--Lagrange equations of Eq.~\eqref{eq:E2}. Similarly to Ref.~\cite{haas2021morphoelasticity}, they are
\begin{subequations}\label{eq:d}
\begin{align}
\dfrac{\mathrm{d}N_s}{\mathrm{d}s}&=\tilde{f}_s\left(\dfrac{N_\phi-N_s}{\tilde{r}}\cos{\tilde{\psi}}+\tilde{\kappa}_sT\right),\label{eq:dNsB}\\
\dfrac{\mathrm{d}M_s}{\mathrm{d}s}&=\tilde{f}_s\left(\dfrac{M_\phi-M_s}{\tilde{r}}\cos{\tilde{\psi}}-T\right),\\
\dfrac{\mathrm{d}T}{\mathrm{d}s}&=-\tilde{f}_s\left(\tilde{\kappa}_sN_s+\tilde{\kappa}_\phi N_\phi+T\dfrac{\cos{\tilde{\psi}}}{\tilde{r}}\right),
\end{align}
\end{subequations}
together with the geometric relations from Eqs.~\eqref{eq:drdz} and~\eqref{eq:dpsi},
\begin{align}
&\dfrac{\mathrm{d}\tilde{r}}{\mathrm{d}s}=\tilde{f}_s\cos{\tilde{\psi}},&&\dfrac{\mathrm{d}\tilde{\psi}}{\mathrm{d}s}=\tilde{f}_s\tilde{\kappa}_s.\label{eq:drB}
\end{align}
In Eqs.~\eqref{eq:d}, 
\begin{subequations}
\begin{align}
N_s&=\dfrac{\alpha_{ss}E_s+\alpha_{s\phi}E_\phi+h\bigl(\beta_{ss}K_s+\beta_{s\phi}K_\phi\bigr)}{\tilde{f}_\phi},\\
N_\phi&=\dfrac{\alpha_{s\phi}E_s+\alpha_{\phi\phi}E_\phi+h\bigl(\beta_{s\phi}K_s+\beta_{\phi\phi}K_\phi\bigr)}{\tilde{f}_s},\\
M_s&=\dfrac{h\bigl(\beta_{ss}E_s+\beta_{s\phi}E_\phi\bigr)+h^2\bigl(\gamma_{ss}K_s+\gamma_{s\phi}K_\phi\bigr)}{\tilde{f}_\phi},
\end{align}

\phantom{.}\vspace{-18pt}
\begin{align}
M_\phi&=\dfrac{h\bigl(\beta_{s\phi}E_\phi+\beta_{\phi\phi}E_\phi\bigr)+h^2\bigl(\gamma_{s\phi}K_s+\gamma_{\phi\phi}K_\phi\bigr)}{\tilde{f}_s},\\
T&=-N_s\tan{\tilde{\psi}}.
\end{align}
\end{subequations}
These equations are closed by the definitions $\tilde{f}_s=1+E_s$, $\tilde{f}_\phi=\tilde{r}/r=1+E_\phi$, $\tilde{\kappa}_\phi=\sin{\tilde{\psi}}/\tilde{r}$, in which the undeformed configuration is a spherical cap of radius $R=1$,
\begin{align}
r(s)=\sin{s}\quad\text{for }0\leq s\leq L.
\end{align}
\subsubsection{Boundary conditions}
We solve these equations subject to the following boundary conditions: At the posterior pole $s=0$, we impose $\tilde{r}=0$ and $\tilde{\psi}=0$. At the phialopore $s=L$, we impose the no-force and no-torque conditions $N_s=0$, $M_s=0$.
\subsubsection{Numerical implementation}
We implement these equations and boundary conditions in \texttt{fortran} for numerical solution in \texttt{auto-07p}~\cite{auto}; the code is available at Ref.~\cite{Note1}. Appendix~B of Ref.~\cite{haas2021morphoelasticity} provides a more general description of the numerical implementation, including of the singularity of Eqs.~\eqref{eq:d} at $s=0$ where $\tilde{r}=0$.
\subsection{Constriction singularity}
The intrinsic volume element in Eq.~\eqref{eq:dV0} is singular at $\zeta^0=1/\kappa^0$. This corresponds to the geometric singularity of perfect cell constriction. Hence $\zeta^0\leq|h^0|/2\leq 1/|\kappa^0|$, so $|H|\leq 1$. Equivalently, $|\eta|\leq 4/3$ from Eq.~\eqref{eq:etas} or
\begin{subequations}
\begin{align}
|\kappa^0|\leq\dfrac{8}{3h}.\label{eq:perfectc}
\end{align}
As $\kappa^0$ approaches this constriction limit, the coefficients in Eqs.~\eqref{eq:coeffs} diverge like $(1-|H|)^{-3}$, so Eq.~\eqref{eq:E2} loses\linebreak asymptoticity. This happens when ${(1-|H|)^{-3}=O\bigl(\varepsilon^{-1}\bigr)}$, i.e., when $\smash{1-|H|\sim a\varepsilon^{1/3}}$, for some ${a>0}$. This corresponds to ${4/3-|\eta|\sim2a \varepsilon^{1/3}}$ from Eq.~\eqref{eq:etas}. With $\varepsilon\sim h/R$, and in our units in which $R=1$, this becomes
\begin{align}
|\kappa^0|\lesssim\dfrac{8}{3h}-\dfrac{4a}{h^{2/3}}.\label{eq:strongc}
\end{align}
\end{subequations}
By comparison with the limit of perfect constriction given by Eq.~\eqref{eq:perfectc}, we shall refer to the limit \eqref{eq:strongc} as the limit of strong constriction. In Figs.~\figrefp{fig3}{b} and~\figrefp{fig3}{c}, we plot the boundary defined by Eq.~\eqref{eq:strongc} for different $a$. 
\subsection{Numerical implementation of the cell shape changes of \emph{P.~californica}}
The intrinsic curvatures defined in Eq.~\eqref{eq:k0} are discontinuous. For numerical reasons, we regularise these discontinuities over a small arclength distance $\Delta s$. We therefore extend Eq.~\eqref{eq:k0} by imposing
\begin{align}
\kappa_s^0(s)=\kappa_\phi^0(s)&=\left\{\begin{array}{ll}
1 &\text{if } s<W-\Delta s,\\
1-(k+1)\dfrac{s-(W-\Delta s)}{2\Delta s}\hspace{-26mm}\phantom{.}&\\
&\text{if }W-\Delta s\leq s<W+\Delta s,\\
-k &\text{if }s>W+\Delta s,\\
\end{array}\right.
\end{align}
in our numerical calculations, for which we arbitrarily took $\Delta s=0.01$. This is much smaller than the single-cell width $\ell\approx 0.3$ that we estimate (\citeappendix{appC}) for \emph{P.~californica} from Fig.~5(a) of Ref.~\cite{hohn2016distinct}. This is therefore in turn consistent with an abrupt change of intrinsic curvature at the front of the wave of cell shape changes.

\begin{figure}[t]
\centering\includegraphics{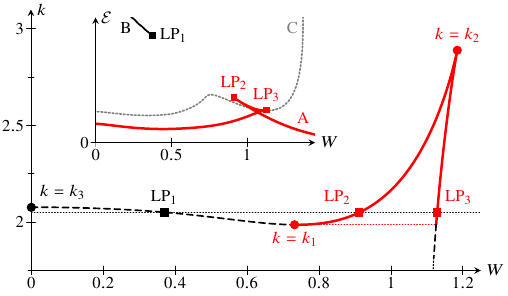}
\caption{Numerical calculation of the critical curvatures $k_1$ and $k_2$. Plot of a branch of fold points in $(W,k)$ space, highlighting bifurcations (disk markers) at $k=k_1<k_3<k_2$ and three fold points $\text{LP}_1,\text{LP}_2,\text{LP}_3$ (square markers) at a fixed value of $k$ (dotted line). Inset: plot of $\mathcal{E}$ against $W$ for this $k$, showing three solution branches A, B, C and the location of $\text{LP}_1,\text{LP}_2,\text{LP}_3$ on these branches. See text for further explanation. Parameter values: $L=1.75$, $h=0.15$, inset: $k=2.05$.}
\label{fig9}
\end{figure}

\section{\uppercase{Bifurcation numerics}}\label{appB}
To determine the critical curvatures $k_1$ and $k_2$ numerically, we computed branches of fold points in $(W,k)$ space~\wholefigref{fig9} by two-parameter continuation in \texttt{auto-07p}~\cite{auto} for fixed values of the remaining model parameters $h,L$. 

At intermediate values of $k>k_1$, three such fold points appear in the $(W,\mathcal{E})$ diagram~\wholefigrefi{fig9}: two fold points on the branch A connected to the undeformed configuration, and another fold point on the branch B, not connected to A, that contains uninverted solutions at $W=0$~\figref{fig2}{b}. We denote these fold points by $\text{LP}_1,\text{LP}_2,\text{LP}_3$ in increasing order of the corresponding value of $W$. Branch A has, moreover, a branch point between $\text{LP}_2$ and $\text{LP}_3$ at which it connects to a branch C of solutions of higher energy that are not relevant physically.\looseness=-1

Thus $\text{LP}_1$ and $\text{LP}_2$ merge at the local minimum of the branch of fold points in \textwholefigref{fig9}. Hence this minimum defines $k=k_1$. For $k<k_1$, but a single fold point remains; this corresponds to the fold on the branch in \textfigref{fig2}{a} containing the inverted solution at $W=0$. Similarly, $\text{LP}_2$ and $\text{LP}_3$ merge with the branch point mentioned above at the local maximum of this branch of fold points~\wholefigref{fig9}. This local maximum is a cusp, and it defines $k=k_2$.

Additionally, $\text{LP}_1$ disappears at $k=k_3>k_1$. At this value, branch B disappears. In other words, the uninverted shape with $W=0$ is not a solution at larger $k$.

\begin{figure}[t]
\centering\includegraphics{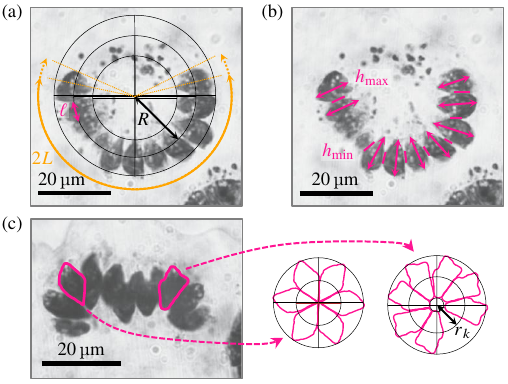}
\caption{Estimation of model parameters. (a)~Estimation of the radius $R$ and the extent $L$ of the cell sheet and of the single-cell width~$\ell$ based on construction of an estimated midplane of the cell sheet. (b)~Estimation of the cell sheet thickness~$h$ as the average of the cell height $h_{\text{max}}$ and the extent $h_\text{min}$ of the cell-cell contact region. (c)~Estimation of the intrinsic curvature of the cell sheet from maximally constricted cells: juxtaposing traced cell outlines estimates the radius of curvature $r_k=1/k$. Experimental light micrographs of \emph{P.~californica} were obtained from Ref.~\cite{hohn2016distinct}. Scale bars: $20\,\text{\textmu m}$.}
\label{fig10}
\end{figure}

\section{\uppercase{Estimation of model parameters}}\label{appC}

In this Appendix, we discuss the estimates of the model parameters reported in Table~\ref{table1}.

First, we estimated the radius $R$ of the cell sheet, its angular extent $L$, and the single-cell width $\ell$~\figref{fig10}{a} from $\mathcal{N}=1$ light micrograph of an uninverted \emph{P.~californica} embryo~\cite{hohn2016distinct}. To this end, we inscribed and circumscribed a circle around the embryo~\figref{fig10}{a}. Their average defined a midline, on which we measured $R$ and $L$. We used the extent of the cell at the phialopore in this micrograph as a proxy for the uncertainty on $L$. We also estimated $\ell=L/m\approx 0.3$, where $m=12$ is the number of cells in the cross section in \textfigref{fig10}{a}.

Next, we estimated the thickness $h$ of the cell sheet by measuring the cell height $h_\text{max}$ and the extent $h_\text{min}$ of the region over which neighbouring cells are in contact~\figref{fig10}{b}, in the same light micrograph of \emph{P.~californica}. The arithmetic mean of these quantities and the quadratic mean of their standard deviations defined our estimate of $h$ and its uncertainty.

Finally, to estimate the intrinsic curvature $k$, we traced outlines of maximally constricted cells~\figref{fig10}{c}. By juxtaposing these cells, we defined a radius of curvature $r_k=1/k$~\figref{fig10}{c}. We analysed $\mathcal{N}=8$ cells in this way; the resulting average and standard deviation defined our estimate of $k$ and its uncertainty.

We obtained estimates of $h$ and $L$ for other species of Volvocaceae [Figs.~\ref{fig5} and \figrefp{fig6}{c}] in an analogous manner.

\section{\uppercase{Volvocaceae and Goniaceae}}\label{appD}
In Table~\ref{table_ref}, we report the numbers $N$ of cells of individuals for all species in the families Volvocaceae and Goniaceae for which we could find an estimate or range of $N$. The data in this table are based on Refs.~\cite{grove1915pleodorina,chodat1931quelques,pocock1955studies,starr1962new,berg1970structure,starr1970volvox,karn1972sexual,akiyama1977illustrations,marchant1977colonymain,ireland1981inversion,nozaki1982morph,nozaki1983morphology,batko1989gonium,nozaki1989morphological,nozaki1989pleodorina,nozaki1990volvulina,nozaki1991pandorina,nozaki1992ultrastructure,nozaki1993asexual,nozaki2001morphology,hallmann2006morphogenesis,nozakimorphology,yamada2008taxonomic,solari2008volvox,hayama2010morphology,iida2011embryogenesis,hohn11,nozaki2011new,isaka2012description,iida2013cleavage,nozaki2014new,nozaki2015morphology,hohn2016distinct,nozaki2017rediscovery,nozaki2018morphology,kimbara2019morphological,nozaki2019morphology,nozaki2019morphologyb,nozaki2022morphology}, most of which we found using the Algaebase database~\cite{algaebase}.

All of the species collected in Table~\ref{table_ref} undergo inversion \cite{hallmann2006morphogenesis}, except for the two species in the genus \emph{Astrephomene}, which form spherical colonies by cell division, without the need for inversion~\cite{nozaki94,yamashita16}.

\begin{table*}[hb!]
\caption{Ranges or estimates of the numbers $N$ of cells of individuals in selected species in the families Volvocaceae and Goniaceae. The table includes all species for which we could find such an estimate. Most of the references cited were found using the AlgaeBase database~\citep{algaebase}.\label{table_ref}}
\begin{ruledtabular}
\begin{tabular}{cccc}
genus\footnote{Genera marked $\dag$ are in the family Goniaceae; all other genera are in the family Volvocaceae.}&species & range or estimate of $N$ & source of the range or estimate of $N$\\
\hline
\multirow{2}{*}{\emph{Astrephomene$^{\dag,}$\footnote{\emph{A.~gubernaculifera} and \emph{A.~perforata} form spherical colonies by cell division, without the need for inversion~\cite{nozaki94,yamashita16}.}}}&\emph{A.~gubernaculifera}&32--64&\citet{nozaki1983morphology}\\
&\emph{A.~perforata}&32--64&\citet{nozaki1983morphology}\\
\hline
\multirow{2}{*}{\emph{Colemanosphaera}}&\emph{C.~angeleri}&16--32&\citet{nozaki2014new}\\
&\emph{C.~charkowiensis}&16--32&\citet{nozaki2014new}\\
\hline
\multirow{4}{*}{\emph{Eudorina}}&\emph{E.~compacta}&16--32&\citet{nozaki2019morphology}\\
&\emph{E.~elegans}&16--32&\citet{marchant1977colonymain}\\
&\emph{E.~minodii}&16--32&\citet{nozaki2001morphology}\\
&\emph{E.~unicocca}&16--32&\citet{yamada2008taxonomic}\\
\hline
\multirow{7}{*}{\emph{Gonium$^\dag$}}&\emph{G.~dispersum}&1 or 16--32&\citet{batko1989gonium}\\
&\emph{G.~maiaprilis}&8--32&\citet{hayama2010morphology}\\
&\emph{G.~multicoccum}&8--32&\citet{pocock1955studies}\\
&\emph{G.~octonarium}&8&\citet{pocock1955studies}\\
&\emph{G.~pectorale}&4--16&\citet{iida2013cleavage}\\
&\emph{G.~quadratum}&8--16&\citet{nozaki1993asexual}\\
&\emph{G.~viridistellatum}&8--16&\citet{nozaki1989morphological}\\
\hline
\multirow{3}{*}{\emph{Pandorina}}&\emph{P.~colemaniae}&8--16&\citet{nozaki1991pandorina}\\
&\emph{P.~minodii}&8--16&\citet{chodat1931quelques}\\
&\emph{P.~morum}&8--16&\citet{hallmann2006morphogenesis}\\
\hline
\emph{Platydorina}&\emph{P.~caudata}&16--32&\citet{iida2011embryogenesis}\\
\hline
\multirow{6}{*}{\emph{Pleodorina}}&\emph{P.~illinoisensis}&16--32&\citet{grove1915pleodorina}\\
&\emph{P.~japonica}&64--128&\citet{nozaki1989pleodorina}\\
&\emph{P.~californica}&64--128&\citet{hohn2016distinct}\\
&\emph{P.~sphaerica}&64--128&\citet{nozaki2017rediscovery}\\
&\emph{P.~starrii}&32--64&\citet{nozakimorphology}\\
&\emph{P.~thompsonii}&16--32&\citet{nozakimorphology}\\
\hline
\multirow{17}{*}{\emph{Volvox}}&\emph{V.~africanus}&1500--6000&\citet{nozaki2022morphology}\\
&\emph{V.~aureus}&1000--3000&\citet{akiyama1977illustrations}\\
&\emph{V.~barberi}&10\,000--50\,000&\citet{solari2008volvox}\\
&\emph{V.~capensis}&2000--6200&\citet{nozaki2015morphology}\\
&\emph{V.~carteri}&1400--3000&\citet{nozaki2018morphology}\\
&\emph{V.~dissipatrix}&7000--23\,000&\citet{nozaki2019morphologyb}\\
&\emph{V.~ferrisii}&5000--8000&\citet{isaka2012description}\\
&\emph{V.~gigas}&1500--3000&\citet{berg1970structure}\\
&\emph{V.~globator}&2000--6000&\citet{hohn11}\\
&\emph{V.~kirkiorum}&3000--6000&\citet{isaka2012description}\\
&\emph{V.~obversus}&2100--4000&\citet{karn1972sexual}\\
&\emph{V.~ovalis}&1000--2000&\citet{nozaki2011new}\\
&\emph{V.~pocockiae}&$\approx$ 1500&\citet{starr1970volvox}\\
&\emph{V.~powersii}&500--1000&\citet{berg1970structure}\\
&\emph{V.~rousseletii}&4700--11\,800&\citet{kimbara2019morphological}\\
&\emph{V.~tertius}&338--613\footnote{By contrast, Ref.~\cite{hallmann2006morphogenesis} states $N\approx 1000$ for \emph{V.~tertius}.}&\citet{ireland1981inversion}\\
&\emph{V.~zeikusii}&4600--13\,000&\citet{nozaki2019morphologyb}\\
\hline
\multirow{3}{*}{\emph{Volvulina}}&\emph{V.~compacta}&8--16&\citet{nozaki1990volvulina}\\
&\emph{V.~pringsheimii}&4--16&\citet{starr1962new}\\
&\emph{V.~steinii}&4--16&\citet{nozaki1982morph}\\
\hline
\emph{Yamagishiella}&\emph{Y.~unicocca}&16--32&\citet{nozaki1992ultrastructure}\\
\end{tabular}
\end{ruledtabular}
\end{table*}
\bibliography{main}
\end{document}